\documentclass[twocolumn,nopacs,preprintnumbers]{revtex4}
\usepackage{graphicx}
\usepackage{bm}
\usepackage{color}
\usepackage[normalem]{ulem}

\usepackage{physics,dsfont}
\usepackage{subfigure}

\graphicspath{%
    {converted_graphics/}
    {/}
}

\begin{document}

\title{Fast and robust quantum state transfer in a topological Su-Schrieffer-Heeger chain with Next-to-Nearest-Neighbour interactions}

\author{Felippo M. D'Angelis$^{1}$, Felipe A. Pinheiro$^{1}$, David Gu\'ery-Odelin$^{2}$, Stefano Longhi$^{3,4}$, Fran\c{c}ois Impens$^{1}$}

\affiliation{$^1$ Instituto de F\'{i}sica, Universidade Federal do Rio de Janeiro,  Rio de Janeiro, RJ 21941-972, Brazil
\\
$^2$ Laboratoire Collisions, Agr\'egats, R\'eactivit\'e, IRSAMC, Universit\'e de Toulouse, CNRS, UPS, France 
\\
$^3$ 
Dipartimento di Fisica, Politecnico di Milano, Piazza L. da Vinci 32, I-20133 Milano, Italy\\
$^4$ 
IFISC (UIB-CSIC), Instituto de Fisica Interdisciplinar y Sistemas Complejos, Palma de Mallorca, E-07122. Spain}

\date{\today}

\begin{abstract}
We suggest a method for fast and robust quantum-state  transfer in a Su-Schrieffer-Heeger (SSH) chain, which exploits the use of next-to-nearest-neighbour (NNN) interactions. The proposed quantum protocol combines a rapid change in the topological edge state, induced by a modulation of nearest-neighbour interactions, with a fine tuning of NNN interactions operating a counter-adiabatic driving which cancels nonadiabatic excitations. We use this shortcut technique on the edge states in order to obtain a quantum state transfer on a single dimerized chain and also through an interface that connects two dimerized Su-Schrieffer-Heeger chains with different topological order. We investigate the robustness of this protocol against both uncorrelated and correlated disorder, and demonstrate its strong resilience to the former. We show that introducing spatial correlations in the disorder increases the robustness of the protocol, widening the range of its applicability. In comparison to traditional adiabatic methods, the short transfer time enabled by the NNN protocol in the SSH chains drastically improves the fidelity of the quantum state transfer.

\end{abstract}

\date{\today}

\maketitle

\section{Introduction}

A promising route to develop quantum information architectures, resilient against decoherence, is to implement the operation on a subset of quantum states that benefit from a natural protection -- for instance when they belong to a specific symmetry class which is immune to decoherence~\cite{DFS03}. The robustness of topological protection~\cite{Kitaev01} has turned topological quantum systems into excellent candidates of quantum computing platforms~\cite{TopologicalNetwork17}, which has motivated their implementation in either photonic~\cite{Kraus2012,Chapman2016,StJean17,Science2018} or atomic systems~\cite{AtomicPlatform1}. In order to obtain universal quantum computation, such  architectures must be able to perform a finite set of elementary tasks with a high degree of  reliability, among which local operations on single qubits, generation of entangled states, and quantum states transfer within the quantum register~\cite{TopologicalNetwork17}. 

In this work, we put forward a fast and resilient protocol for quantum state transfer in a Su-Schrieffer-Heeger (SSH) chain~\cite{SSH}, which provides the simplest one-dimensional lattice with topologically-protected edge states. Among relevant previous examples of transfer protocols, we mention the use of  time-independent  fields  in coupled quantum dots~\cite{QuantumDots04}, protocols relying on single-qubit Rabi flopping protocols~\cite{RabiFlopping1,TopologicalNetwork17,Estarellas}, the transfer of doublons across a spin chain~\cite{RabiDoublonTransfer}, and Thouless pumping \cite{Kraus2012,Chapman2016}. A strong motivation to develop quantum computation on a topological register is the intrinsic resilience against stochastic perturbations. In this line, a key feature of toplogical quantum state transfer protocols should be to preserve this robustness, and to be resilient against possible imperfections in the dynamical control parameters as well as against local and long-range parasitic fluctuations affecting the spin chain. For this purpose, adiabatic protocols have been considered in 
in single-dimerized spin chains with an even~\cite{Longhi19a} or odd~\cite{Mei2018} number of sites, as well as in SSH chains involving two segment of different topologies~\cite{Longhi19b}. These protocols operate on a subset of edge states, which eigen-energies are distant from the band, and rely on well-known adiabatic techniques such as Landau-Zener(LZ)~\cite{Longhi19a} or Stimulated Raman Adiabatic Passage~(STIRAP)~\cite{Longhi19b} transitions. These procedures have shown a good resilience against uncorrelated disorder, but their speed is intrinsically limited by the adiabaticity requirement - both to avoid unwanted nonadiabatic transition within the edge-state multiplicity, and from the edge-state multiplicity to the energy bands. Fast quantum state transfer is also a desirable feature to limit the impact of decoherence. Thus, the application of shortcut-to-adiabaticity techniques~\cite{RMPShortcut}, suggested to accelerate  state transfer protocols in topologically-trivial systems \cite{uff1,uff2,uff3,uff4,uff5}, seems suitable in the context of topological chains. Initial steps have been taken in this direction~\cite{TopologicalShortcut1,TopologicalShortcut2}, with few experimental realizations so far~\cite{TopologicalShortcutExp1,TopologicalShortcutExp2}.

Here, we present an accelerated protocol of high performance on both a single SSH chain and a succession of two SSH chains having distinct topological orders. Our approach is based on an engineering of Next-to-Nearest Neighbor(NNN) interactions between the sites of the chain, which exactly cancels the excitations towards the band even for a strongly accelerated transfer. Our method is extremely robust against disorder, both uncorrelated and correlated. In this respect, it outperforms previous adiabatic protocols for single dimerized SSH chains~\cite{Mei2018}. Interestingly, our method is also more robust than the STIRAP protocol in SSH chains with a topological interface~\cite{Longhi19b}.

The paper is organized as follows. In Section II, we present the procedure and illustrate the method on a single SSH chain.  In Section III, we apply our method to a SSH chain with two distinct topologies and obtain a fast quantum state transfer across the topological interface. In Section IV, we investigate numerically the resilience of the excitation transfer protocol to both diagonal and off-diagonal static disorder -either spatially uncorrelated or correlated - and to fluctuations of the driving NNN interaction strengths. Finally, the main conclusions are outlined in Sec.V.\\

\section{Principle of the Next-to-Nearest-Neighbour interaction assisted transfer}

\begin{figure}[htbp]
    \centering
    \includegraphics[width=8.5cm]{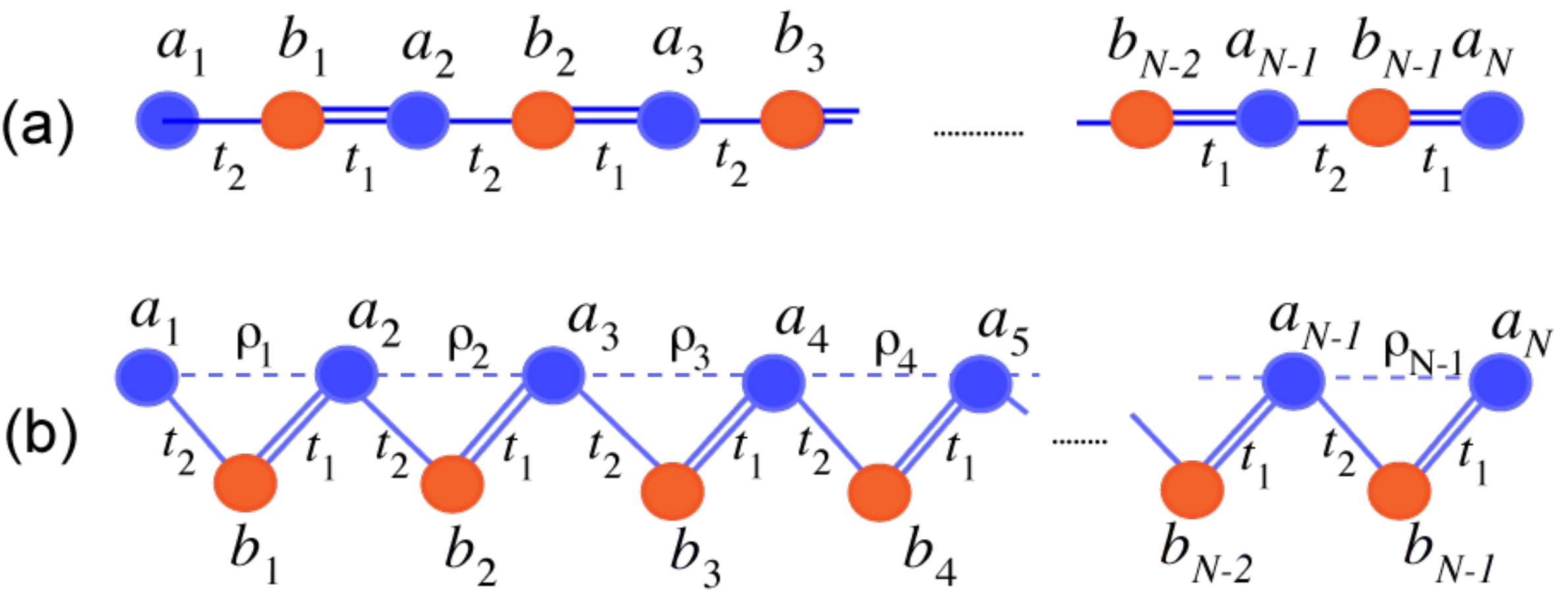}
    \caption{\label{fig:scheme} Diagram showing the SSH chain with (a) nearest neighbour hopping and (b) additional next-to-nearest neighbour hopping (dashed line).}
\end{figure}

 A standard model of quantum state transfer exploiting topological edge states is provided by a SSH chain formed by two sub-lattices with chiral symmetry \cite{TopologicalNetwork17, Estarellas, Longhi19a,Mei2018}. Several experimental platforms have already enabled the emulation of SSH chains: in nanophotonics with polaritons in an array of micropillars \cite{StJean17}, in granular chains \cite{granular}, in a 1D bichromatic lattice \cite{atala}, in arrays of optically trapped 
Rydberg atoms \cite{AtomicPlatform1}  or in a momentum-space lattice \cite{meier}.  
 
 To exemplify our method, we consider in this section a single chain that contains respectively $N$ sites in the sublattice $A$ and $N-1$ sites in the sublattice $B$, with alternate nearest-neighbour (NN) time-dependent hopping amplitudes $t_1(t)$ and $t_2(t)$ (see  Fig.~\ref{fig:scheme}a). This latter requirement can be experimentally implemented. For instance, in Ref.~\cite{meier}, the effective tunneling transition are field driven and as such can be readily made time-dependent.  

In the basis of Wannier states and in the single excitation sector, this quantum system is described by the SSH Hamiltonian
\begin{equation}
\hat{H}_0(t)= \sum_{n=1}^{N-1} \left( t_{2}(t) | B_n \rangle \langle A_n| +  t_{1}(t) | A_{n+1} \rangle \langle B_n|  \frac {} {} \!  \right)+   {\rm h. c.} 
\label{eq1}
\end{equation}
The quantum state of the chain is represented by a $2N-1$ state vector $ | \psi (t) \rangle$. For an odd number of sites, the SSH Hamiltonian has a single zero-energy mode lying in the gap separating the two energy bands, and localized at the left edge of the chain for $t_2 \ll t_1$:
\begin{eqnarray}
\label{eq:eigenvector}
| \phi_0 \rangle & = & \mathcal{N} \sum_{n=1}^{N} \: \left( \frac {-t_2} {t_1} \right)^{n-1} | A_n \rangle 
\label{eq2}
\end{eqnarray} 
This zero-energy mode is parametrized  according to $\epsilon = - t_2/t_1$, which determines the normalization constant $ \mathcal{N}= [(\epsilon^2 - 1)/(\epsilon^{2N}-1)]^{1/2}$.  At $t_1=t_2$, i.e. at the band closing point, this
state is fully delocalized along the chain. Finally, for $t_2 \gg t_1$, the eigenstate $| \phi_0 \rangle$ is localized at the right end of the chain.
Let us assume that the state vector is initially at the left chain boundary, taking for instance $| \psi (0) \rangle = | A_1 \rangle$. A full transfer to the right end of the chain amounts to obtaining  $|\langle A_{N}| \psi(T) \rangle|^2=1$ at the final time $T$ of the protocol.
 By slowly changing the time-dependent hopping parameters $t_1(t)$ and $t_2(t)$ so that $\epsilon(t)$ evolve from $\epsilon(t=0) \ll1$ to 
$\epsilon(t=T) \gg 1$, one can induce such a transfer by an adiabatic evolution of the instantaneous Hamiltonian eigenstate $|\phi_0(t) \rangle$. However, in order to prevent nonadiabatic transitions to the continuum of states, according to the adiabatic theorem this strategy requires --roughly speaking- that the temporal variation of the hopping terms be much slower than the energy gap between the edge mode and the band. Since in the adiabatic evolution the gap closes at the time for which $t_1=t_2$, nonadiabatic effects might provide a severe limitation to the quantum state transfer efficiency.\\ Here we unveil a powerful means to overcome the adiabaticity  constraint with a dynamical control of NNN interactions inspired by counter-diabatic driving methods.
 We assume from now on that the SSH chain is dressed by NNN couplings which can be controlled dynamically, as 	shown in Fig.~\ref{fig:scheme}(b). The quantum chain is thus governed by the following Hamiltonian
\begin{equation}
\hat{H}(t)= \hat{H}_0(t)+ \sum_{n=1}^{N-1} \left(  i \rho_n(t)  | A_{n+1} \rangle \langle A_n|  +   {\rm H. c.}  \frac {} {}  \!  \right) \, .
\label{hamilto}
\end{equation}
 The $\pi/2$ gauge phase between NN and NNN couplings is introduced so that the inverse-engineering procedure yields at any time real-valued hopping amplitudes $\rho_n(t)$, as shown below. The purpose of the time-dependent NNN hoppings $\rho_n(t)$ is to cancel exactly non-adiabatic transitions from the time-dependent eigenvector $|\phi_0(t) \rangle$. Differently from the traditional counter-adiabatic driving method~\cite{Berry09} that prevent non-adiabatic transitions from all instantaneous eigenstates, our approach cancels only the adiabatic transitions from the topological mode $| \phi_0(t) \rangle$, which is sufficient to achieve a reliable transfer. This trick considerably simplifies the form of the driving terms, enabling one to implement the method with only NNN interactions  in one of the sublattices.

Note that the Hamiltonian~\eqref{hamilto} does not present the chiral symmetry usually associated with the topological protection of adiabatic quantum state transfers in spin chains~\cite{TopologicalNetwork17}. The reason for this is twofold. First, the SSH chain has an odd number of sites. Second, the NNN couplings connect sites of the same sublattice and thus break the chiral symmetry. In the adiabatic limit, the NNN hopping terms cancel and the Hamiltonian~(\ref{hamilto}) boils down to the tight-binding Hamiltonian~\eqref{eq1}. The non-invertibility of the latter then ensures robustness against off-diagonal disorder of the zero-energy eigenstate~\cite{Yuce19} despite the absence of chiral symmetry. For nonadiabatic quantum state transfers, as discussed later in Section IV, the resilience of the NNN protocol comes rather from the short transfer durations which mitigate the influence of disorder.

 From the Schr\"odinger equation with $\hbar=1$ associated to Eq.~(\ref{hamilto}), we get the following set of coupled equations that describe the time evolution of the probability amplitudes $a_n(t)$ and $b_n(t)$ in each sublattice:
\begin{align}
	i \dv{a_n}{t} &= t_2 b_n + t_1 b_{n-1} - i \rho_n a_{n+1} + i \rho_{n-1} a_{n-1},   \label{eq:new_an} \\
	i  \dv{b_n}{t} &= t_2 a_n + t_1 a_{n+1}, \label{eq:new_bn} 
\end{align}
with $2 \leq n \leq N-2$ and where explicit time-dependence have been omitted to simplify notations. The amplitude probabilities associated to the chain boundaries satisfy the specific equations:
\begin{align}
	i  \dv{a_1}{t} & = t_2 b_1 - i \rho_1 a_{2},    \label{eq:a1} \\
	i \dv{a_{N}}{t} &= t_1 b_{N-1} + i \rho_{N-1} a_{N-1}.  \label{eq:aN_1}
\end{align}
 Our goal is to inverse-engineer these equations for a given time-dependence of the NN hopping terms $t_{1,2}(t)$  such that the time-dependent $| \phi_0(t) \rangle$ state, defined by Eq.(\ref{eq2}), is an {\em exact} solution to the time-dependent Schr\"odinger equation. To determine the profiles of the NNN hopping amplitudes $\rho_n(t)$, let us first notice that, as this state vector has no overlap with ``B'' sites, one has $b_n(t)=0$ for $n=1,...,N-1$. Furthermore, the form of the wave-vector $|\phi_0(t) \rangle$ guarantees that Eqs.\eqref{eq:new_bn}  are fulfilled by construction for $b_n(t)=0$, independently of the specific choices for the hopping amplitudes $t_{1,2}(t)$ and $\rho_n(t)$. Incidentally, this makes the protocol immune to the eventual presence of additional parasitic NNN couplings between the ``B'' sites. Equation~\eqref{eq:a1} determines $\rho_1(t)$, while Eqs.~\eqref{eq:new_an} fix the remaining NNN hopping terms $\rho_n(t)$ for $n=2,...,N-1$. Preservation of the norm of the quantum state vector guarantees that Eq.~\eqref{eq:aN_1} is fulfilled. The wave-vector $| \phi_0(t) \rangle$ evolves indeed on a manifold of dimension $N-1$, so that any quantum trajectory associated to a unitary evolution can be obtained with an appropriate parametrization of the $N-1$ hopping amplitudes $\rho_n(t)$.  From Eqs.(\ref{eq:new_an}) and (\ref{eq:a1}), one ends up with the following set of recurrence relations:
\begin{eqnarray}
\rho_1 &= & - \frac {1} {a_2} \dv{a_1}{t},  \label{eq:rho1}\\
\rho_n  &= & \frac {1} {\epsilon^2} \rho_{n-1}  - \frac {(n-1) \hbar} {\epsilon^2} \dv{\epsilon}{t} - \frac {1} {\epsilon} \dv{\ln(\mathcal{N})}{t},     \,  \label{eq:rhon} 
\end{eqnarray}
with $2 \leq n \leq N-1$.

It is instructive to consider the quantum-state transfer in a $3$-sites chain, which corresponds to the well-known STIRAP scheme. In this case our method reduces to the counterdiabatic driving method~\cite{david2010}, involving a single coupling $\rho_1(t)$. The Schr\"odinger equation yields a simple three-equation system for the amplitudes $a_1(t)$, $b_2(t)$ and $a_2(t)$. The relevant adiabatic zero-energy mode is given by 	$a_n = \epsilon^{n-1}/\sqrt{1+\epsilon^2} $ for $n=1,2$ and the recurrence relation~\eqref{eq:rho1} yields the NNN coupling $\rho_1(t) =  (1+\epsilon^2)^{-1} \dv{\epsilon}{t}.$ Hopping amplitudes $t_1(t),t_2(t)$ must fulfil the boundary conditions $|t_1(0)/t_2(0)|\ll 1$ and $|t_1(T)/t_2(T)|\gg 1$ required to achieve a high fidelity in the quantum state transfer.
There is an infinity of possible choices, and one may privilege a particular profile depending on the feasibility of the experimental implementation. By taking for instance a sinusoidal modulation of the NN couplings $t_1(t) = t_0 \cos [ \pi t/ (2T) ]$ and $t_2(t) = t_0 \sin [\pi t / (2T)]$ with $t_0$ an arbitrary constant energy, one finds a particularly simple, constant counter-driving term $\rho_1(t) = -\pi  /2T$. As expected, this counter-adiabatic coupling vanishes in the adiabatic limit. The expression of the NNN hopping terms $\rho_n(t)$ becomes increasingly intricate as we consider longer chains.

\begin{figure}[htbp]
    \centering
    \includegraphics[width=8.5cm]{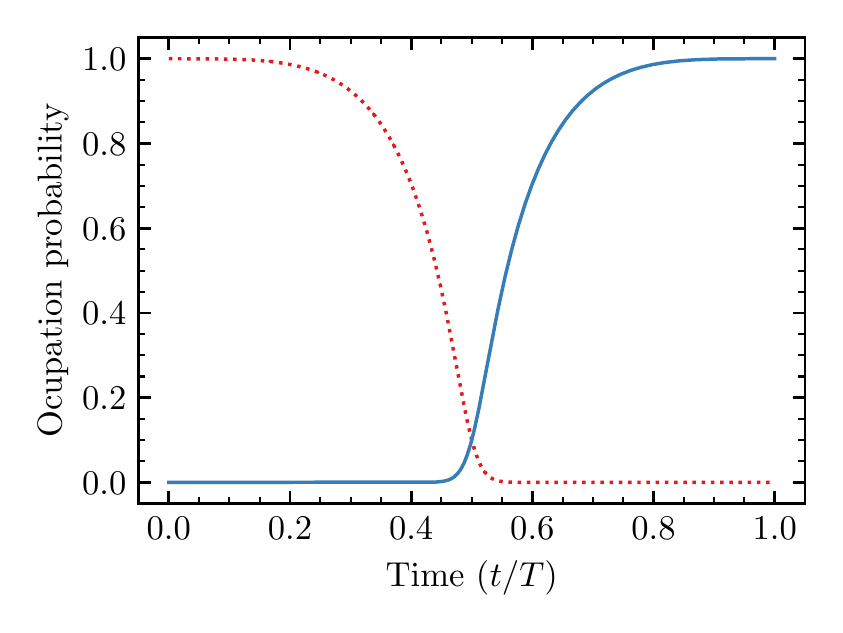}
    \caption{\label{fig:1} Time-dependent probability of occupation of the sites $A_1$ (red-dotted line) and $A_{N}$ (blue solid line) in a SSH chain with NNN-assisted coupling. The chain comprises $N=10$ sites in sublattice $A$. NN hopping amplitudes used in the numerical simulations are $t_1(t)=t_0 P(t/T)$ and $t_2(t)=t_0[1-P(t/T)]$ with the polynomial $P(x)=2 x^3-3x^2+1$ and with a total transfer time $T = 2 / t_0$.}
\end{figure}

\begin{figure}[htbp]
    \centering
    \includegraphics[width=8.5cm]{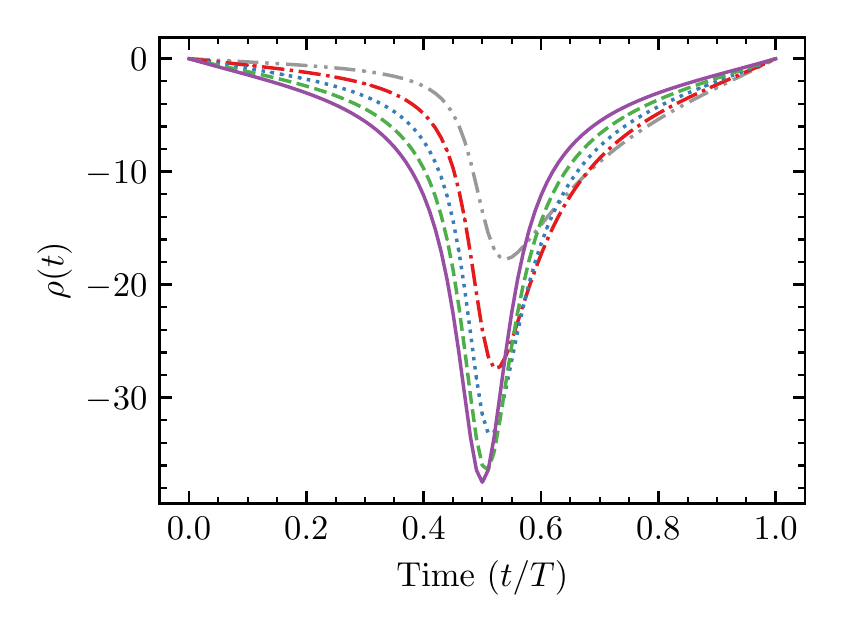}
    \caption{\label{fig:2} Profile of the NNN couplings $\rho_n(t)$ (in units of $t_0$) as prescribed by the NNN-interaction assisted transfer for $n=1,..,5$ (dash-dot-dotted gray, dash-dotted red, dotted blue, dashed green and solid purple lines respectively). Parameter values are as in Fig.\ref{fig:1}.} 
\end{figure}

To illustrate the validity of our procedure when considering longer chains, we apply the method to a SSH chain of $2N-1=19$ atoms in the configuration depicted in Fig \ref{fig:scheme}(b) for a total duration $T=2/ t_0$, where $t_0$ is the largest value taken by the hopping amplitude $t_1$ (or $t_2$) in the transfer process. Clearly, such a short time transfer violates the adiabaticity criterion and, without counter-diabatic driving,
 the transfer efficiency is greatly degraded (less than 1 \%). In order to speed-up numerical calculations, and similarly to many shortcut-to-adiabaticity protocols~\cite{RMPShortcut,DavidSciRep17,Impens17,Impens20}, we have parametrized the quantum state trajectory with polynomial -- the simplest functions to fulfill the required set of boundary conditions.
Precisely, we have taken NN hoppings $t_1(t)=t_0P(t/T)$ and $t_2(t)=t_0(1-P(t/T))$ with the polynomial $P(x)=2 x^3-3x^2+1$. $P(x)$ is indeed the  lowest order polynomial fulfilling the conditions $P(0) = 1$, $P(1) = 0$ together with a cancellation of the derivatives at the boundary of the time interval. The latter is necessary to ensure a smooth parabolic switching on/off of the NN hoppings. The corresponding results, depicted in Fig.\ref{fig:1}, have been obtained from a full numerical resolution of the Schr\"odinger equation in presence of the time-dependent NN and NNN interactions. They confirm the consistency of our approach, i.e. that a perfect transfer is achieved thanks to the counter-diabatic driving with NNN interactions.
Figure \ref{fig:2} shows the time-dependence of the NNN coefficients $\rho_n(t)$ along the chain. Symmetry considerations show that $\rho_{N-1-n}(t)=\rho_n(T-t)$, so that we only represent the couplings $\rho_n(t)$ for $n=1,...,\lceil{N/2}  \rceil $.

\section{Next-to-Nearest-Neighbour-interaction-assisted transfer across a topological interface}

\begin{figure}[h!]
	\centering
	\includegraphics[width=9 cm]{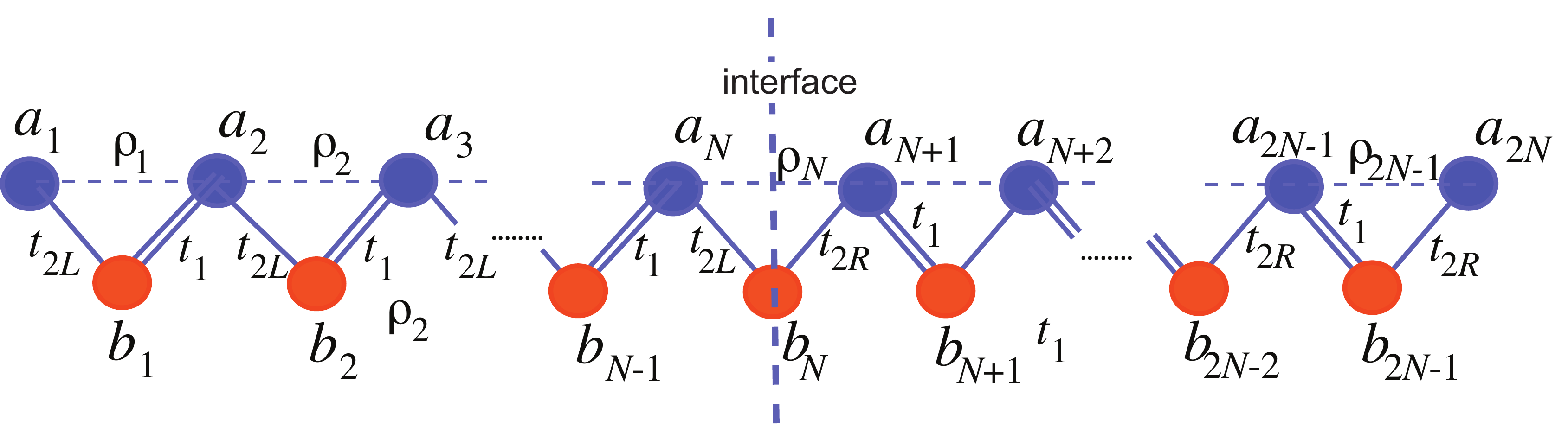} 
	\caption{\label{fig:8} SSH chain with a topological interface, dressed with dynamically controlled NNN interactions.}
\end{figure}

We extend here the fast NNN transfer protocol to an SSH chain presenting a topological interface, i.e. involving two segments of different topologies. As discussed hereafter, this method strongly outperforms adiabatic transfers considered so far in these systems~\cite{Longhi19b}. We consider a SSH chain of $4N-1$ atoms involving two fragments of different topologies as represented in Fig.~\ref{fig:8}.
The single excitation sector Hamiltonian now reads:
 \begin{eqnarray}
 \label{eq12}
& \hat{H}_1(t) & \! \! =  t_{2 L}(t) \!  \sum_{n=1}^{N} \: | B_n \rangle \langle A_n| +  t_{1}(t) \! \sum_{n=1}^{N-1} | A_{n+1} \rangle \langle B_n|   \\ \! \!  & +  & \! \! \! \! \! \! t_{2 R}(t) \! \! \sum_{n=N}^{2N-1} | A_{n+1} \rangle \langle B_n| + t_{1}(t) \! \! \! \! \sum_{n=N+1}^{2N-1} | A_{n} \rangle \langle B_n|+ \!  {\rm h. c.} \nonumber 
\end{eqnarray}
The topology on each side of the interface depends on the parameters $\epsilon_{L} = - t_{2 {L}} / t_1$ and $\epsilon_{R} = - t_{2  {R}} / t_1$. When $|\epsilon_L|$ and $|\epsilon_R|$ are both greater than unity, or both smaller than unity, the sides of the chain have different topologies, and the chain exhibits a topological interface at the site $b_N$. From now on, we consider specifically the regime for which $0< |\epsilon_L| < 1$ and $0< |\epsilon_R| < 1$. In this case the chain has three topological modes isolated from the energy bands. In the limit $\epsilon_{L,R} \ll 1,$  two of these modes are localized at the chain boundaries, while the third zero-energy mode lies in the vicinity of the interface. A procedure for a quantum state transfer across such a topological SSH chain has been proposed in Ref.~\cite{Longhi19b}. This approach relies on a STIRAP protocol within the 3-state multiplicity of topological states. A dark state, built up from a superposition of two edge states, is then evolved adiabatically from one extremity of the chain to the other. This evolution is obtained by a variation of the hopping amplitudes $t_{2 \: L,R}(t)$. Although efficient in the adiabatic limit, the bottleneck of this method is the minimum transfer duration imposed by the adiabaticity condition, which rapidly grows with the chain length.

Following the strategy exposed in the previous Sections,  we set up an appropriate engineering of NNN interactions to go beyond this limitation.  We prescribe a trajectory along the following zero-energy mode:
\begin{equation}
| \psi_0  \rangle =\frac {1} {\mathcal{N}} \left( \begin{array}  {c} \epsilon_R^N  \\ 0 \\  \epsilon_L \epsilon_R^N \\ 0 \\... \\ 0  \\ \epsilon_L^{N-1}\epsilon_R^N\\ 0 \\  - \epsilon_L^{N} \epsilon_R^{N-1}   \\ 0 \\ ...\\ 0 \\ - \epsilon_L^{N}  \epsilon_R  \\  0 \\ - \epsilon_L^{N}  \end{array}  \\ \right)  \label{eq:Interface quantum state}
\end{equation}
where we assume from now on $\epsilon_{L,R} \neq 0$. As previously, this wave-vector is restricted to the  sublattice $A$ (odd components). The wave-vector components in the immediate vicinity of the interface are $a_N=\epsilon_L^{N-1}\epsilon_R^N$ and $a_{N+1}=- \epsilon_L^{N} \epsilon_R^{N-1}$. The time dependence of the parameters $ \epsilon_L, \epsilon_R$ has been omitted to simplify notations. $\mathcal{N}$ is a normalization constant setting the norm of the wave-vector to unity, which can be expressed as  $\mathcal{N}= \left[ \epsilon_R^{2N}(1- \epsilon_L^{2N})/(1-\epsilon_L^2)+ \epsilon_L^{2N}(1- \epsilon_R^{2N})/(1-\epsilon_R^2) \right]^{1/2}$ if $\epsilon_{L,R} \neq   1 $.

 The prescribed quantum state $| \psi_0(t) \rangle$, given by ~\eqref{eq:Interface quantum state}, fulfills, by construction, the projection of the Schr\"odinger equation on the sublattice $B$ for any time-dependent NN hoppings. In contrast, this wave-vector only satisfies the  Schr\"odinger equation on the sublattice $A$ if appropriate NNN couplings $\rho_n(t)$ are applied. The Hamiltonian including these NNN couplings reads $\hat{H}(t)=\hat{H}_1(t)+\left(\sum_{n=1}^{2N-1} \: i \rho_n(t)  | A_{n+1} \rangle \langle A_n|+{\rm H.c} \right).$  The NNN couplings are obtained by an inverse engineering of the Schr\"odinger equation.  One obtains the following recurrence relations:
\begin{eqnarray}
\rho_1 &= &- \frac {1} {\epsilon_L} \frac {\dot{a}_1} {a_1},  \nonumber  \\
\rho_{m+1} & = & \frac {\rho_m}  {\epsilon_L^2} - m \frac {\dot{\epsilon}_L} {\epsilon_L^2}+\rho_1, \nonumber \\
\rho_N & = & - \frac {\epsilon_R} {\epsilon_L^2} \rho_{N-1}+(N-1) \frac {\epsilon_R \dot{\epsilon}_L} {\epsilon_L^2}+ \frac {\epsilon_R} {\epsilon_L} \frac {\dot{a}_1} {a_1}, \label{eq:recurrence_interface} \\
\rho_{N+1} & = & - \frac {\epsilon_R^2} {\epsilon_L^2} \rho_N - N \frac {\epsilon_R \dot{\epsilon}_L} {\epsilon_L}- \epsilon_R  \frac {\dot{a}_1} {a_1}, \nonumber \\
\rho_{n+1} & = & \epsilon_R^2 \rho_n - N \frac {\epsilon_R \dot{\epsilon}_L} {\epsilon_L}  + (n-N) \dot{\epsilon}_R  -  \dot{\epsilon}_R  \frac {\dot{a}_1} {a_1},  \nonumber
\end{eqnarray}
where the $m$ and $n$ indices belong to the intervals $1 \leq m\leq N-2$ and $N+1 \leq n \leq 2 N-2$ and with the first wave-vector component $a_1= \mathcal{N}^{-1} \epsilon_R^{N}$. Note that the couplings in the immediate vicinity of the interface satisfy different equations. One has again $2N$ equations to be satisfied by $2N-1$ independent parameters $\rho_n(t).$ Thanks to the conservation of the wave-vector norm, this system is not overdetermined.\\

\begin{figure}[htpb]
\begin{subfigure}
  \centering
  \includegraphics[width=8.5cm]{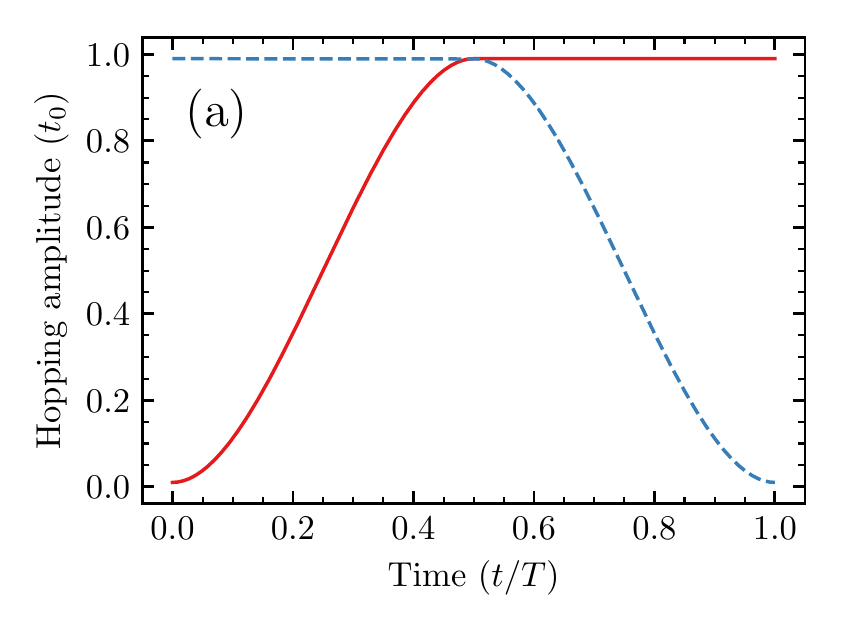}  
\end{subfigure}\\
\begin{subfigure}
  \centering
  \includegraphics[width=8.5cm]{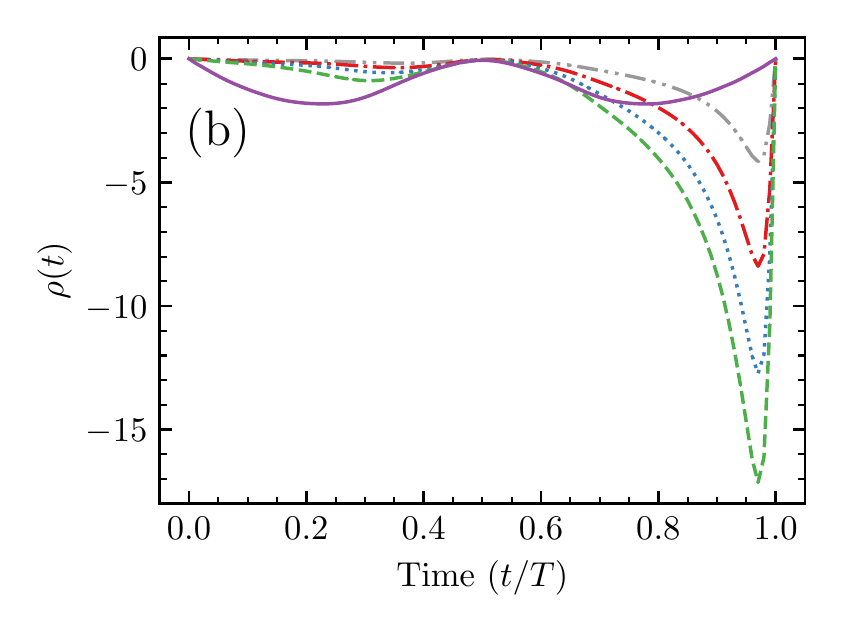}  
\end{subfigure}\\
\begin{subfigure}
 	\centering
  	\includegraphics[width=8cm]{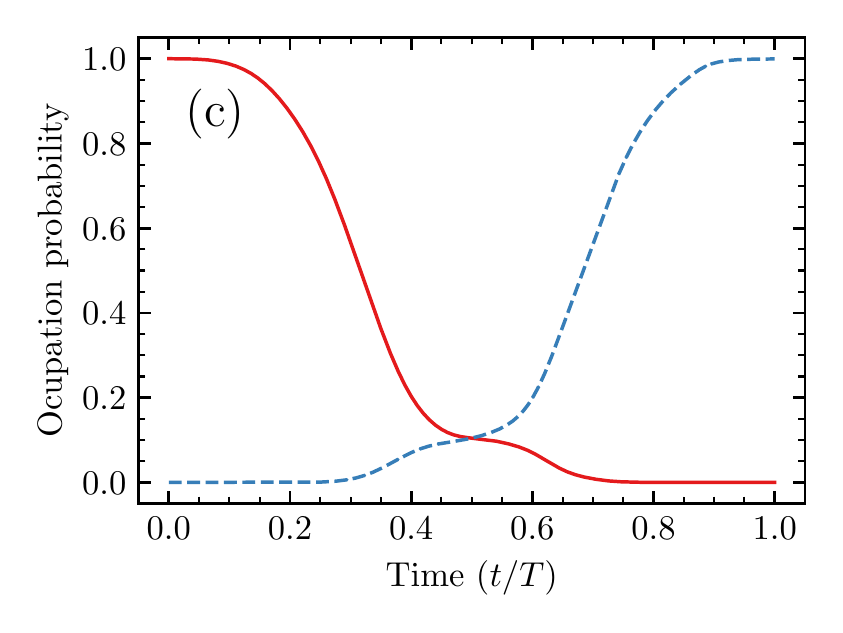}  
	\end{subfigure}
    \caption{\label{fig:interfaceCouplings} (a) Hopping amplitudes $t_{2R}(t)$ and $t_{2L}(t)$ as a function of the normalized time $t/T$ (solid blue line and dashed red line respectively). (b) Couplings $\rho_n(t)$ (in units of $t_0$)  for $n=1,...,5$ as a function of time $t/T$ (dash-dot-dotted gray, dash-dotted red, dotted blue, dashed green and solid purple lines respectively). We have taken $N=5$, corresponding to a SSH chain of $4N-1=19$ sites, and a total time $T=40 /t_0$. (c) Time-dependent probability of occupation of the sites $|A_1\rangle$ (red line) and $|A_{2N}\rangle$ (dashed blue line) in a SSH chain with topological interface and NNN-assisted transfer protocol. The chain comprises $4N-1=19$ sites, and the total duration is $T=40 /t_0$. }
\end{figure}

The choice of the hopping amplitudes $t_1(t)$, $t_{2 L}(t)$ and $t_{2 R}(t)$ must be compatible with a transfer of the wave-vector~(\ref{eq:Interface quantum state}) from the left to the right boundary of the chain. The initial location at the left end is achieved if $|\epsilon_L(0)| \ll 1$ and a finite value of $|\epsilon_R(0)| < 1,$ while the location at the right end is obtained at the final time $T$ if $|\epsilon_R(T)| \ll 1$ with a finite $|\epsilon_L(T)| < 1$. From now on, we simply set $t_1(t)=t_0$ and take for the hopping amplitudes: $t_{2 L}(t)=t_0 f(2 t /T)$ for $t<T/2$, $t_{2 L}(t)=t_0 f(T/2)$ for $t \geq T/2$, and $t_{2 R}(t)=t_{2 L}(T-t) $. $f(x)$ is a polynomial defined as $f(x)= \delta + (1- 2 \delta)(3x^2 -2x^3)$~\cite{footnote1}, with $\delta =0.01$, and the corresponding amplitudes are depicted on Fig.~\ref{fig:interfaceCouplings}(a). By inspection of Eq.~\eqref{eq:Interface quantum state}, this choice provides initial and final locations at the left and right extremities of the chain respectively. One also notes that  $0<| \epsilon_{L}(t)|< 1$ and $0<| \epsilon_{R}(t)|< 1$ at any time, so that the chain maintains a topological interface through the whole process. At the half time $t=T/2$, one has $\epsilon_{L}(T/2)=\epsilon_{R}(T/2)=0.99,$ corresponding to a state delocalized on the whole chain for the considered values of $N$.

The set of recurrence relations~(\ref{eq:recurrence_interface}) may induce NNN coupling amplitudes $\rho_n(t)$ scaling exponentially with $n$, or even divergent coupling strengths for specific times. This would be a serious drawback, as the implementation of such couplings would then require an unrealistic amount of energy in long topological chains. A suitable profile for the NN hoppings should thus yield counterdiabatic driving terms of finite amplitude. As shown in Fig.~\ref{fig:interfaceCouplings}(b), our choice of NN hopping amplitudes $t_{2 L,R}(t)$ yields smooth and well-behaved NNN couplings $\rho_n(t)$. These NNN couplings vanish at the half time $T/2$, when the time-derivatives of the NN hoppings cancel. At this specific time, the time derivative of the wave-vector~\eqref{eq:Interface quantum state} cancels and thus the counterdiabatic NNN couplings are not needed. Figure~\ref{fig:interfaceCouplings}(c) displays the temporal profile of the occupation probabilites associated to the chain boundaries for a total time $T=40/t_0$. As a consistency check, one notes that a perfect population transfer is obtained at the final time.

\section{Resilience of the transfer protocol against correlated/uncorrelated disorder}

Any realistic implementation of the protocol will involve a finite amount of disorder, either in the control field operating the NNN couplings, either in the realization of the SSH chain itself. 
We thus discuss here the resilience of the counter-diabatic protocol to a simplification of its execution, and to the presence of disorder in the realization of the protocol or in the implementation of the SSH chain. The presence of stochastic fluctuations, either in the on-site energies or in off-diagonal elements of the Hamiltonian, may decrease the efficiency of the quantum state transfer. Nevertheless, the NNN-interaction assisted transfer presents a surprising resilience against both correlated and uncorrelated disorder, outperforming in this respect previous schemes using adiabatic methods~\cite{Mei2018,Longhi19b}. Correlated disorder is known to play an important role in various transport phenomena, for both classical~\cite{conley2014} and quantum waves~\cite{li2011}, and it is thus relevant to investigate if it constitutes a limitation in the implementation of transfers via shortcut to adiabaticity. In the following we note $N_A$ the total number of sites in the sublattice $A$ (with our conventions $N_A=N$ for the single-dimerized SSH chain, and $N_A=2N$ for the SSH chain with a topological interface).  As one seeks to obtain a perfect population transfer  to the right-end of the chain, the quantum fidelity $\mathcal{F}$ of the protocol corresponds to the transfer probability $p_{A_{N_A}}=| \langle A_{N_A} | \psi(T) \rangle|^2$.

\subsection{Simplification and resilience to an imperfect realization of the protocol in a simple SSH chain}

For a SSH chain of  $2N-1$ atoms, our transfer procedure involves the accurate control of $N-1$ couplings associated to NNN interactions with distinct time dependence.  From this respect, the accurate implementation of these different couplings, such as those shown in Fig.~\ref{fig:2}, seems challenging experimentally. To partially overcome this limitation, we suggest here an alternative and simpler implementation. Figure~\ref{fig:2} reveals that the shape of the couplings $\rho_n(t)$ depends less on the index $n$ when $n \to N/2$, i.e. when this index  corresponds to sites near the middle of the chain. In contrast, the NNN couplings $\rho_n(t)$ and $\rho_{n+1}(t)$ have quite different profiles if the index $n$ is chosen near the edges. With these considerations in mind, we introduce a simplified approximate protocol where most of the NNN coefficients are assigned a common time-dependent profile. Specifically, we set $\tilde{\rho}^{(i)}_n(t)=\rho_{N/2}(t)$ for $n=i,...,N-i$, and $\tilde{\rho}^{(i)}_n(t)=\rho_n(t)$ for $n=1,..i-1$ and $n=N-i+1,..N-1$. The NNN couplings $\rho_n(t)$ correspond to the exact NNN transfer protocol. The lower the index $i$, the simpler the protocol $\tilde{\rho}^{(i)}_n(t)$. The protocol obtained for $i=1$ involves only a single function for all NNN couplings, while the choice $i = \lceil N/2 \rceil$ amounts to the exact procedure. In this alternative implementation, the trade-off between the quality of the quantum state transfer and the simplification brought by the substitution is crucial. Remarkably, values of $i$ of a few units already yield a quantum fidelity close to unity. Table \ref{tab:approx_protocol} gives the transfer probability achieved for the approximate protocol $\tilde{\rho}^{(i)}_n(t)$ in a chain with $N_A=10$ sites in sublattice $A$ and for several approximation levels $i$, demonstrating the fast convergence of the method.

\begin{table}[htpb]
	\caption{\label{tab:approx_protocol} Transfer probability $p_{\rm A_{10}}$ versus index $i$ for the simplified NNN protocol in a single SSH chain (Fig.~\ref{fig:scheme}). The index $i$ measures the approximation degree of the exact NNN protocol.}
	\begin{ruledtabular}
		\begin{tabular}{l c c c c}
			\hline			
    		i & 1 & 2 & 3 & 4 \\
    		\hline
  			$p_{\rm A_{10}}$ & 0.12 & 0.88 & 0.87 & 0.99\\
  			\hline  
 		\end{tabular}
	\end{ruledtabular}
\end{table}

\begin{figure}[h!]
	\centering
	\includegraphics[width=8.5cm]{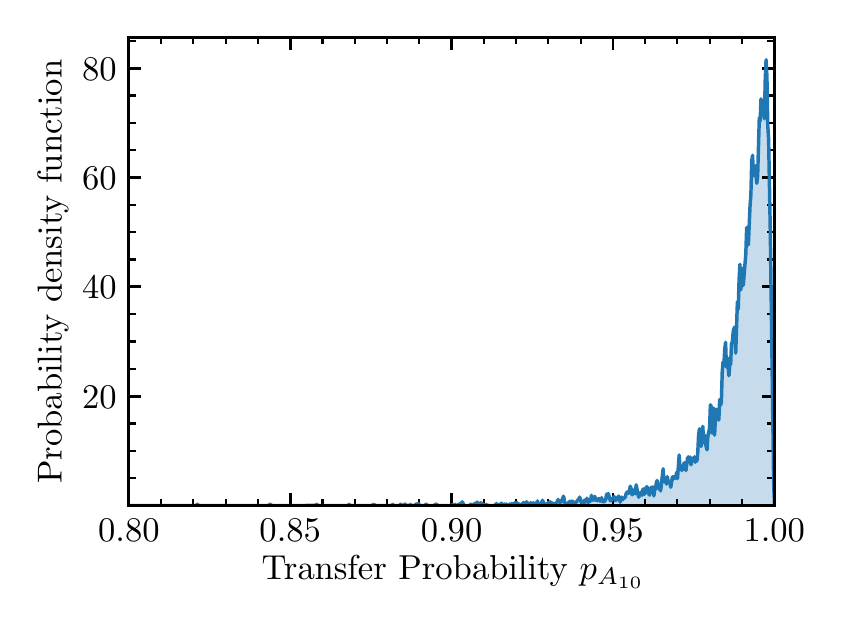}
	\caption{\label{fig:5} Resilience to an imperfect realization of the  NNN couplings: Probability density function of the transfer probability $p_{A_{N_A}}=|\langle A_{N_A} |\psi(T) \rangle |^2$  obtained from $10.000$ realizations of the NNN protocol in a single SSH chain  (see Fig.~\ref{fig:scheme}) with $N_A=10$ sites in the sublattice $A$.  We have used common stochastic bias factor multiplying the NNN couplings $\rho_n(t)$, centered on unity (error-free case), and which follows a Gaussian distribution $G(\alpha)$ of unit standard deviation and area $\alpha = 0.2$. The NN hopping amplitudes, the total duration and the error-free NNN couplings $\rho_n(t)$ are identical to Fig.~\ref{fig:1}.}
\end{figure}

We now discuss the robustness of our protocol against stochastic fluctuations of the counter-diabating driving terms. For this purpose, we consider a noisy realization of the complete protocol (i.e.~without the approximation proposed above) in the simple SSH chain, where the NNN couplings now correspond to stochastic variables $\hat{\rho}_n(t).$ We introduce a random variable $G(\alpha)$, following a centered Gaussian distribution (i.e.~null expectation value) with unit standard deviation and area $\alpha$, and set $\hat{\rho}_n(t)= (1+G(\alpha))\rho_n(t)$.   The amplitude of the NNN couplings are multiplied by a common time-independent stochastic factor, which could be related for instance to a bias in the control laser field. 
Figure \ref{fig:5} shows the distribution of the transfer probability $p_{A_{N_A}} $ (treated from now on as a random variable) obtained from the resolution of the Schr\"odinger equation for $10.000$ realizations of the stochastic NNN couplings. One finds that the transfer probability $p_{A_{N_A}} $ remains above $95\%$ for almost all the realizations despite the presence of significant fluctuations in the NNN couplings (error of the order of $20 \%$). Most of the realizations show indeed a transfer probability above $98\%$.

\subsection{Robustness of the NNN protocol against static uncorrelated disorder in the SSH chain}

We discuss here the resilience of the protocol against static uncorrelated disorder. This disorder may either affect the onsite energies of the SSH chain (diagonal disorder), or alter nondiagonal terms such as the NN couplings $(t_1,t_2)$. The considered NNN protocol is designed for a SSH chain with an odd number of sites. To highlight the benefits of our method, we shall compare it to well-known adiabatic protocols operating on SSH chains with an identical number of sites, and thus with the same parity.  In the single-dimerized SSH chain, we compare the performance of the NNN scheme with the adiabatic protocol of Ref.~\cite{Mei2018}. Regarding the SSH chain with a topological interface, we compare the robustness of our NNN protocol to the STIRAP-based transfer presented in Ref.~\cite{Longhi19b}. In both systems, our NNN protocol significantly outperforms the adiabatic transfers.

\begin{figure}[htbp]
\begin{subfigure}
  \centering
  \includegraphics[width=8.5cm]{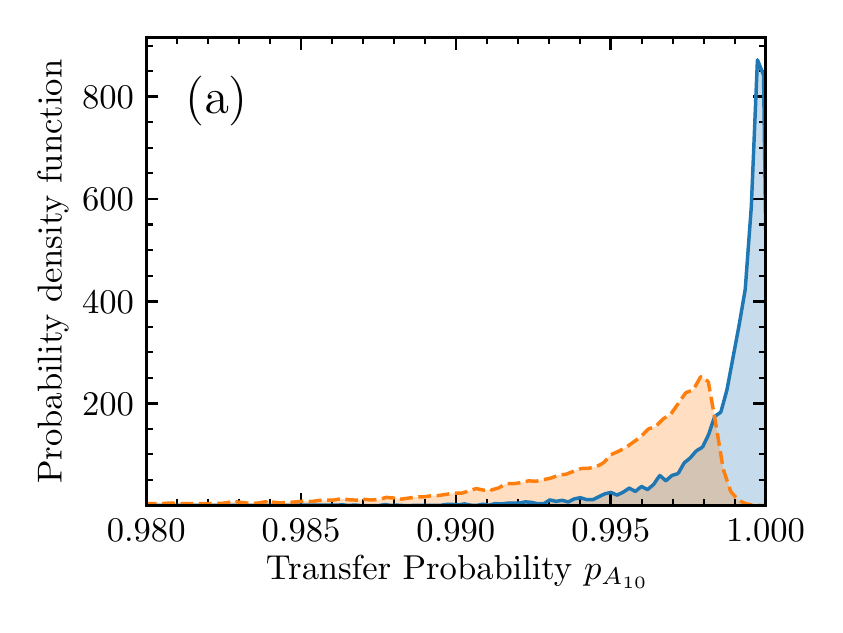}  
\end{subfigure}\\
\begin{subfigure}
  	\centering
  	\includegraphics[width=8.5cm]{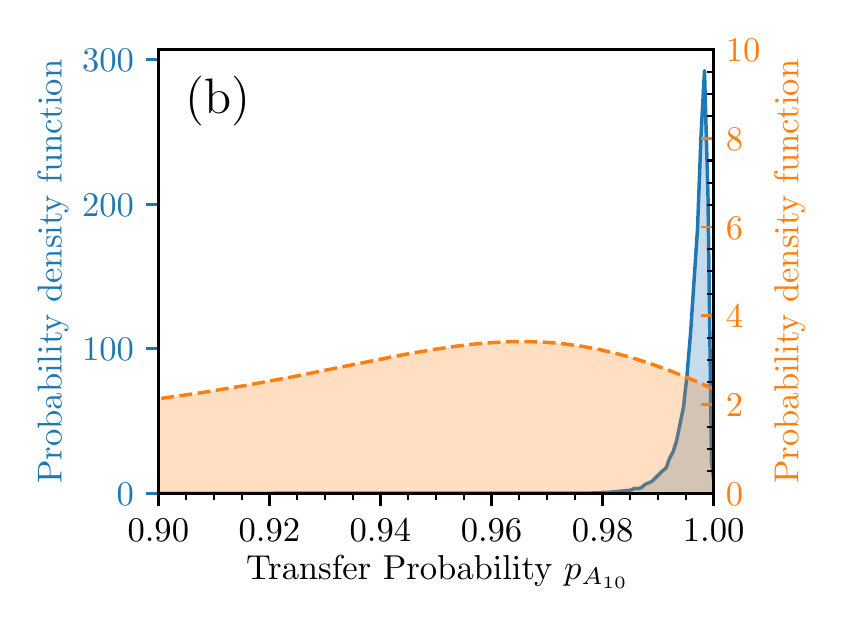}  
	\end{subfigure}
	 \caption{\label{fig:4}Resilience to diagonal disorder: Probability density function of the transfer probability $p_{A_{N_A}}$ for the nonadiabatic NNN protocol (a) in a simple SSH chain (solid blue lines) along with  the distribution associated to the adiabatic protocol (dashed orange line) presented in Ref.~\cite{Mei2018} and (b) in a SSH chain with topological interface (solid blue lines, left scale) compared with the adiabatic STIRAP protocol from Ref.~\cite{Longhi19b} (dashed orange line, right scale). Diagonal terms of the Hamiltonian are subjected to an uncorrelated Gaussian disorder $\delta E= t_0 G(\alpha)$ with $\alpha = 0.1$ for all protocols. Each distribution has been obtained with $10.000$ realizations of the disorder. Parameters of the NNN protocol in the simple SSH chain (a) and  SSH chain with topological interface (b) are identical to Fig~\ref{fig:1} and Fig.~\ref{fig:interfaceCouplings} respectively, except for the transfer time set to $T=2/t_0$. All SSH chains have $N_A=10$ atoms in the $A$ sublattice. The parameters for the adiabatic protocols in (a) and (b) have been extracted from Refs.\cite{Mei2018,Longhi19b} and reproduced in Appendix A for convenience.}

\end{figure}

Let us first investigate the resilience to diagonal Gaussian disorder. Precisely, we assume that the Hamiltonian can be written as $H'(t) = H(t) + \delta H$ with a stochastic contribution $\delta H = {t_0 \: \rm Diag}[G(\alpha)]$ corresponding to a diagonal matrix whose diagonal elements are static, statistically independent random variables following a Gaussian distribution $G(\alpha)$ centred around zero with an area $\alpha$ and unit standard deviation. Note that the sites in the sublattices $A$ and $B$ are equally affected by this disorder.  Figure~\ref{fig:4} compares the behavior of the NNN-interaction protocol with adiabatic transfers for both SSH chains (with and without interface), and for the Gaussian diagonal disorder of same amplitude $\alpha=0.1$. For convenience, a description of the considered adiabatic protocols can be found in Appendix A. The duration of the NNN protocol has been set to $T=2/t_0$, while we have considered larger transfer times for the adiabatic protocols in order to fulfill the adiabaticity conditions. The presence of a topological interface in the SSH chain slightly reduces the resilience of the NNN protocol, but its robustness in this context is still impressive.   In both the single-dimerized and the two-segment SSH chains, the NNN transfer protocol outperforms the adiabatic protocols.

\begin{figure}[htbp]
	\centering
	\includegraphics[width=8.5cm]{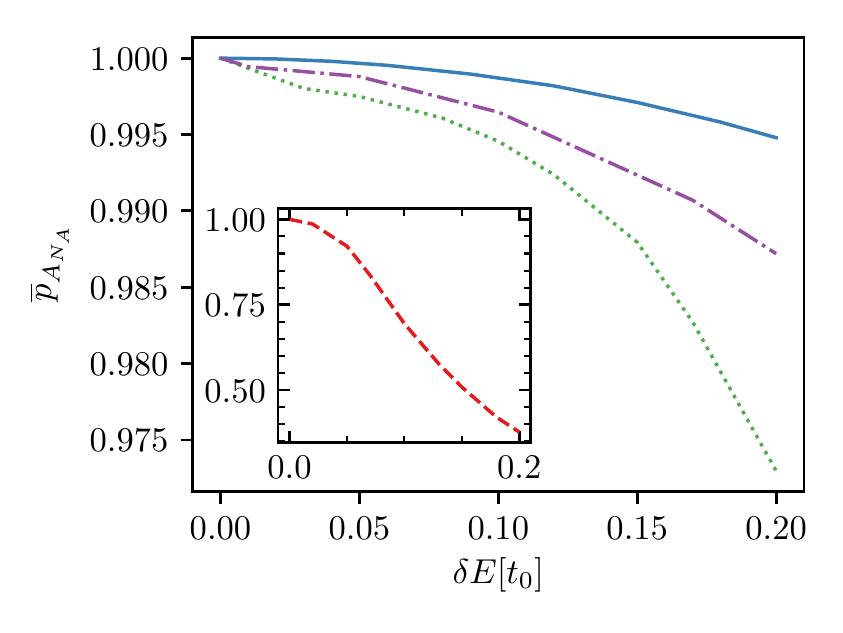} 
	\caption{\label{fig:6}  Resilience to diagonal disorder: Average transfer probability $\overline{p}_{A_{N_A}}$ for the NNN (solid blue line) and adiabatic (dotted green line) protocols in a simple SSH chain, NNN protocol in a SSH chain with topological interface (dash-dotted purple line) and STIRAP protocol (inset, dashed red line) as a function of the diagonal disorder strength $\delta E=t_0 \alpha$ (in units of $t_0$). The NNN and adiabatic protocols have identical parameters as in Fig.\ref{fig:4}. The diagonal disorder has a similar form as in Fig.\ref{fig:4} with a strength $\delta E$ taken in the considered range. Each value of $\overline{p}_{A_{N_A}}$ has been obtained through an averaging over 10.000 realizations.}
\end{figure}

In order to obtain a more quantitative picture of the robustness of the NNN protocol against diagonal disorder, we investigate the average transfer probability  $\overline{p}_{A_{N_A}}$ versus the disorder amplitude $\delta E=t_0 \alpha$. Figure \ref{fig:6} sums up the results obtained for the NNN transfer in the single and two-fragment SSH chains, as well as for the two adiabatic protocols of Refs.~\cite{Mei2018,Longhi19b}.  As expected, the average transfer probabilities decrease with the strength of disorder, indicating that none of these protocols benefits from the presence of uncorrelated disorder. Nevertheless, Fig.~\ref{fig:6} shows that, in the two types of SSH chains, the NNN transfer clearly outpferforms the considered adiabatic protocols over this range of disorder amplitudes. Indeed, for the disorder strength $\delta E =0.2 t_0$, the NNN protocol yields an average transfer probablity $\overline{p}_{A_{N_A}} \simeq 99.5 \%,$ while the adiabatic protocol only reaches a fidelity of $\overline{p}_{A_{N_A}} \simeq 97 \%.$ The advantage is even more drastic in the two-segment SSH chain: for a disorder strength $\delta E =0.1 t_0,$ the NNN protocol reaches an average probabilty $\overline{p}_{A_{N_A}} \simeq 99.6 \%$ against a probability of $\overline{p}_{A_{N_A}} \simeq 70 \%$ for the STIRAP protocol.

\begin{figure}[htbp]
  	\includegraphics[width=8.5cm]{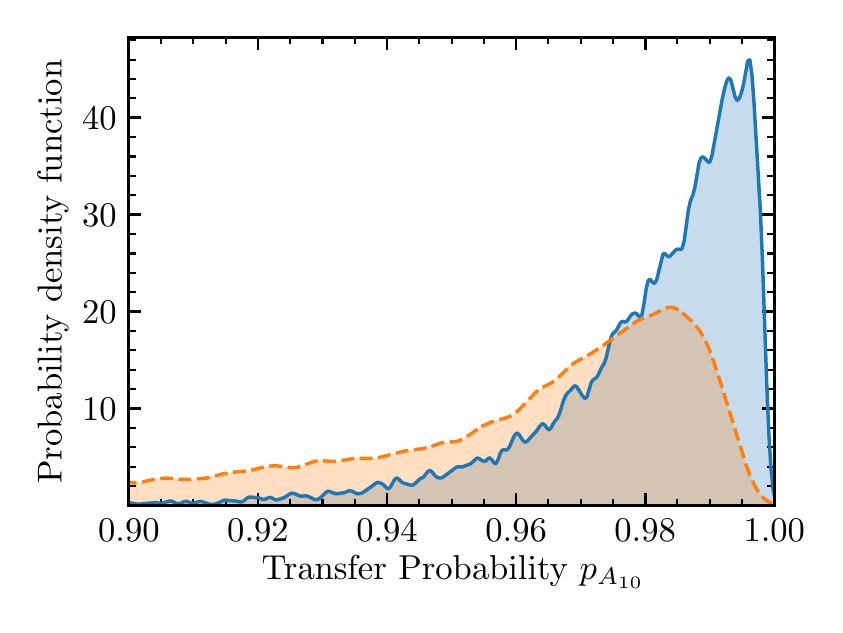}  
	 \caption{\label{fig:OffDiagDisorder} Resilience against off-diagonal disorder: Probability density function of the transfer probability $p_{A_{N_A}}$ for the NNN protocols in a simple SSH chain (solid blue line) and in a SSH chain with topological interface (dotted orange line) for nondiagonal disorder in the off-diagonal terms corresponding to $t_1$, $t_{2}$. The distribution of transfer probabilties has been obtained through $10.000$ realizations of the disorder in a single-dimerized SSH chain and in a SSH chain with topological interface. The NNN protocol and the SSH chain length are the same as in Fig.~\ref{fig:4}.}	 
\end{figure}

The advantage of the NNN protocol over the adiabatic procedures is even more impressive in presence of nondiagonal disorder in the Hamiltonian. We consider specifically local random fluctuations in the NN hoppings along the SSH chains. We set for each realization $H'(t) = H(t) + \delta H$
with $\delta H= \sum_{n=1}^{N_A-1} \delta t_{1 n}  | A_{n+1} \rangle \langle B_n|  +  \sum_{n=1}^{N_A-1} \delta t_{2 n}  | B_{n} \rangle \langle A_n| + {\rm h. c.}$, where the stochastic amplitudes $\delta t_{1n}$ and $\delta t_{2 n}$ are independent identically distributed random variables following the same Gaussian distribution $\delta E=t_0 G(\alpha)$ with $\alpha=0.1$. While the adiabatic protocols yield a very small transfer probability in presence of this disorder, the NNN protocol still achieves average transfer probabilities $\overline{p}_{A_{N_A}} \simeq 98.0 \% $ and $\overline{p}_{A_{N_A}}  \simeq 94.5 \%$ for the single-dimerized and two-segments SSH chains respectively (See Fig.~\ref{fig:OffDiagDisorder}).

\subsection{Robustness against spatially correlated disorder}

In addition to uncorrelated Gaussian disorder, we also investigated the influence of correlated disorder of on-site energies (diagonal disorder) on the fidelity of the transfer protocols for the SSH chain. More precisely, we considered the following power-law correlation \cite{almeida2018}
\begin{equation}
	\delta E_n = \sum_{k=1}^{N_A} \frac{t_0 \alpha}{k^{\gamma/2}}\cos{\left(\frac{4\pi k n}{2 N_A-1}+\phi_k \right)}
\label{eq:powerlawcorrelation}
\end{equation}
where $n$ indicates the site position in the lattice (as far as the disorder is concerned the sites $A$ and $B$ are treated indifferently), $\phi_k$ is a random phase and $\gamma$ is a parameter that sets the correlation length. $\gamma = 0$ corresponds to an uncorrelated disorder while $\gamma>0$ accounts for a disorder with a long-range correlation in the chain. The correlation length is an increasing function of this exponent.

\begin{figure}[htbp]
	\includegraphics[width=8.5 cm]{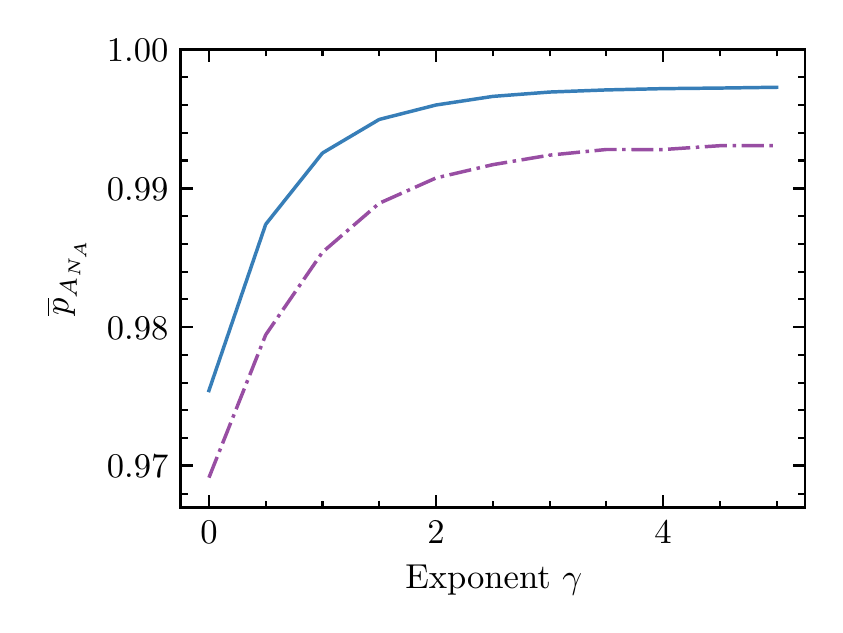} 
	\caption{\label{fig:10} Resilience against spatially correlated disorder: Average transfer probability $\overline{p}_{A_{N_A}}$ at the right boundary as a function of the exponent $\gamma$ for the NNN protocol in a single SSH chain (solid blue line) and in a double SSH chain with a topological interface (dashed-dotted purple line). We have taken a diagonal correlated disorder following the power law~\eqref{eq:powerlawcorrelation} with $\alpha=0.2$, a transfer time $T=2/t_0.$ Each value of  $\overline{p}_{A_{N_A}}$ has been obtained through an averaging over $2.000$ realizations.}
\end{figure}

We have investigated the influence of these correlations in protocols applied in both the single SSH chain and in the SSH chain with a topological interface. Typical numerical results are shown in Fig.~\ref{fig:10}. For both kind of chains, the average transfer probability increases  for higher values of the exponent $\gamma$.  The error rate in the transfer, on the order or $2\%$ and $3\%$ for the uncorrelated disorder in the simple-dimerized and two-fragments SSH chain respectively, can be respectively lowered to $0.3 \% $ and $0.7\%$ by increasing this exponent. 
We conclude that the NNN protocol is more robust against correlated disorder than against uncorrelated Gaussian disorder: increasing the degree of correlation in the disorder improves the performance in the transfer. This finding unveils how disorder, and in particular the emerging concept of disorder engineering by introducing spatial correlations~\cite{conley2014}, may not only degrade but also be used to control transfer protocols. 

\subsection{Physical interpretation of the resilience of the NNN protocol against disorder}
	 
As stated above, our NNN protocol shows a remarkable resilience against both diagonal and off-diagonal disorder. This robustness is not related to topological protection, as chiral symmetry is broken by the NNN terms. It rather stems from the short time scales enabled by nonadiabatic transfers. Indeed, the effects of disorder accumulate over time during the unitary evolution of the quantum system, and thus these effects are less relevant in fast NNN transfers than in slow adiabatic protocols. Accelerating quantum protocols in order to reduce the effect of parasitic couplings is indeed the main motivation for shortcut-to-adiabaticity methods.

The interplay between duration and resilience to disorder is most clearly illustrated by the Anderson localization phenomenon. Consider an excitation initially localized in a site of a lattice
with on-site disorder. At the initial stage of the dynamics one observes a near ballistic spreading of excitation in the lattice (like in the disorder-free lattice). This is so until the diffusion times becomes comparable to the Heisenberg time, which is the longest time an excitation can propagate in the system~\cite{Thouless72,AndersonReview}. At subsequent times, the excitation does not spread anymore and Anderson localization sets in. In order to achieve Anderson localization induced by disorder, it is thus necessary for the system to diffusively evolve for a certain time scale. This diffusion time decreases as the disorder strength increases.

With this argument in mind, one can easily understand why our fast protocol is less sensitive to disorder than any other adiabatic protocol. A given amount of disorder requires a minimum time-scale in order to affect quantum propagation. Conversely, by shortening the total duration of the propagation, one increases the threshold of disorder above which the propagation is significantly altered. This is the key to the robustness of shortcut-to-adiabaticity methods. Provided that the total transfer time $T$ is short enough, and the disorder strength not too large, the impact of some static perturbations in the Hamiltonians~(\ref{eq1},\ref{eq12}) just produces a small effect on the final state $|\psi (T) \rangle$, regardless of the type of disorder (diagonal or off-diagonal). On the other hand, in all adiabatic protocols the transfer time $T$ turns out to be much longer, so that the same strength of disorder affects the final state $|\psi ( T) \rangle$ in a much more drastic way, resulting is a strongly degraded quantum fidelity.

\section{Conclusions and perspectives}

Summing up, we have presented an excitation transfer protocol in SSH chains based on a fast evolution of a topological eigenstate, balanced by a counterdiabatic driving through a dynamical control of NNN interactions. The hopping terms are dynamically controlled in the spirit of current experimental platforms for SSH chains~\cite{meier}. A simplified scheme has been proposed to use a common control field in a large part of the chain. This protocol, which rests on a dynamical control of NN and NNN interactions, shows a strong resilience against fluctuations in the amplitude of the control field and against uncorrelated disorder in the on-site energies or in the off-diagonal elements of the SSH Hamiltonian. Interestingly, the method can be extended to realize fast and robust excitation transfer even in SSH chains displaying two segments of different topologies. This result  constitutes, to our knowledge, the first example of shortcut-to-adiabaticity protocol involving the crossing of a topological interface. In both the single-dimerized and in the two-segments SSH chain, the NNN protocol strongly outperforms usual adiabatic transfer methods. This enhanced resilience stems from the short time transfer enabled by the counterdiabatic driving, which mitigate the influence of disorder on the transfer. Finally, we have found that the presence of correlation in the disorder mitigates the degradation of the fidelity of the quantum state transfer, indicating that judicious engineering of correlated disorder could enhance the efficiency of the transfer protocols in topological chains.

\section*{ACKNOWLEDGMENTS}

This work is part of the INCT-IQ from CNPq.  F.M.D.A., F.A.P. and F.I. acknowledge support from the agencies CNPq
(Grants Universal Faixa B  No. 403366/2016­0 and No. 409994/2018-9), CAPES, and FAPERJ. This work is also funded by the Agence Nationale de la Recherche through Grant No. ANR-18-CE30-0013 (D.G.-O.). S.L. acknowledges the Spanish State Research Agency, through the Severo Ochoa and Mar\'ia de Maeztu Program for Centers and Units of Excellence in R$\&$D (Grant No. MDM-2017-0711).

\section*{APPENDIX A: ADIABATIC PROTOCOLS}

We briefly describe here the adiabatic protocols considered in Section IV. 

Regarding the adiabatic protocol of Ref.~\cite{Mei2018} adressing single-dimerizeed SSH chains with an odd number of sites, we have used the following time-dependent profile for NN hopping terms $t_1(t),t_2(t) $:
\begin{align}
	t_1(t) &= t_0(1 + \cos{\Omega t}) \: \: ({\rm A}1) \nonumber \\
	t_2(t) &= t_0(1 - \cos{\Omega t})  \: \: ({\rm A}2) \nonumber
\end{align}
which correspond to the parameters used in Fig.2 of Ref.~\cite{Mei2018}. In this protocol, the quantum state transfer is  achieved through a displacement of the zero-energy mode along the chain. The final time $T$ must be chosen such as $T = \pi / \Omega $ in order to obtain a displacement from one extremity to the other of the SSH chain. The adiabaticity condition can be written as $\sqrt{t_0 \Omega} \ll \Delta E,$ where $\Delta E$ is the energy gap between the zero-energy mode and the bulk modes. This can be achieved by taking $\Omega = 0.01 t_0$.

 In the STIRAP protocol of Ref.~\cite{Longhi19b} applied to a SSH chain with topological interface, the Hamiltonian~\eqref{eq12} has three nearly-degenerate modes, noted $|L\rangle$, $|R\rangle$ and $|C\rangle$. For long SSH chains, these modes become orthogonal and are localized at the left end, right end and at the topological interface of the chain respectively, and read:
\begin{align}
\nonumber
|L\rangle= N_L\begin{pmatrix}
1 \\
0 \\
X \\
0 \\
X^2 \\
0 \\
\vdots
\end{pmatrix},
|R \rangle = N_R\begin{pmatrix}
\vdots \\
0 \\
Y^2 \\
0 \\
Y \\
0 \\
1
\end{pmatrix},
|C \rangle = N_C\begin{pmatrix}
\vdots \\
X^2\\
0 \\
X \\
1 \\
Y \\
0 \\
Y^2\\
\vdots
\end{pmatrix} \\
 \: \: ({\rm A}3) \nonumber
\end{align}
where we have set $X = -t_{2L}/t_1$ and $Y = -t_{2R}/t_1.$  $N_{L,R,C}$ are appropriate normalization factors. By considering the restriction of the Hamiltonian to the subspace defined by these three modes, the Schr\"odinger equation takes the following form:
\begin{equation}
	i\frac{\rm{d}}{\rm{d}t}|\psi\rangle = \begin{pmatrix}
0 & \Omega_L & 0 \\
\Omega_L & 0 & \Omega_R \\
0 & \Omega_R & 0 
\end{pmatrix} |\psi \rangle
\nonumber  \: \: ({\rm A}4)
\end{equation}
where the state vector $|\psi (t)\rangle$ has been projected on the subspace generated by the three states $\{ |L(t)\rangle , |R(t)\rangle , |C(t)\rangle \}$. This is the basic equation for a STIRAP transfer in a three-level system. The effective Rabi pulsations $\Omega_L(t),\Omega_R(t)$ in~({\rm A}4) are determined by the NN hopping amplitudes $t_1(t)$, $t_{2L}(t)$ and $t_{2R}(t)$.\\

The strategy is to approximate the full SSH chain Hamiltonian~\eqref{eq12} by this reduced Hamiltonian, tailoring the NN hopping coefficients $t_1(t)$, $t_{2L}(t)$ and $t_{2R}(t)$
 in order to achieve a full STIRAP transfer between the states $| L \rangle$ and $| R \rangle$. In order for this approximation to be valid, the transfer time must be long enough as to ensure the adiabiticity condition. We have used the same parameters as in Fig.3 of Ref.\cite{Longhi19b}:  
 $t_1(t) = t_0$, $t_{2L}(t) = t_0 \Omega_m e^{-(t-\delta/2)^2/w^2}$ and $t_{2R}(t) =t_0 \Omega_m e^{-(t+\delta/2)^2/w^2}$. We have set $t_i=-T/2$ and $t_f= T/2$ the initial and final times for the evolution of the NN hopping terms, and set $\delta = w/(3 t_0)$, $w = 3T/16$ and $\Omega_m = 0.9$. The transfer time $T = 900 / t_0$ differs slightly from that used in Ref.\cite{Longhi19b} and has been chosen in accordance with the length of the SSH chain.


\begin{thebibliography}{1}
\bibitem{footnote1} The polynomials considered here are identical to the polynomials of Section II if one takes $\delta=0$. These polynomials yield divergent couplings for the topological SSH chain. Taking a small but finite $\delta$ such that $0< \delta \ll 1$ avoids this divergence at the expense of a negligible loss in fidelity.
\end{thebibliography}
\end{document}